\documentclass[preprint]{aastex6}

\shorttitle{Origin of double RC in the MW Bulge}
\shortauthors{Joo, Lee, \& Chung}

\begin{document}

\title{NEW INSIGHT ON THE ORIGIN OF THE DOUBLE RED CLUMP IN THE MILKY WAY BULGE}

\author{
Seok-Joo Joo\altaffilmark{1,2}, Young-Wook Lee\altaffilmark{2}, and Chul Chung\altaffilmark{2}}

\altaffiltext{1}{Korea Astronomy and Space Science Institute, Daejeon 34055, Korea; sjjoo@kasi.re.kr}
\altaffiltext{2}{Center for Galaxy Evolution Research and Department of Astronomy, Yonsei University, Seoul 03722, Korea;
ywlee2@yonsei.ac.kr, chung@galaxy.yonsei.ac.kr}

\begin{abstract}

The double red clump (RC) observed in the Milky Way bulge is widely interpreted as evidence for an X-shaped structure.
We have recently suggested, however, an alternative interpretation based on the multiple population phenomenon,
where the bright RC is from helium enhanced second-generation stars (G2), while the faint RC is representing
first-generation stars (G1) with normal helium abundance. Here our RC models are constructed in a large parameter space
to see the effects of metallicity, age, and helium abundance on the double RC feature. Our models show that the luminosity
of RC stars is mainly affected by helium abundance, while the RC color is primarily affected by metallicity.
The effect of age is relatively small, unless it is older than 12 Gyr or much younger than 6 Gyr.
The observed double RC feature can therefore be reproduced in a relatively large parameter space,
once $\Delta$Y between G2 and G1 is assumed to be greater than $\sim$0.10. We further show that the longitude dependence
of the double RC feature at $b \approx -8\degr$, which was pointed out by \citet{gon15} as a potential problem of our model,
is well explained in our scenario by a classical bulge embedded in a tilted bar.

\end{abstract}


\defcitealias{lee15}{Paper~I}

\advance\textheight -24pt

\section{INTRODUCTION}

Some years ago, the presence of two red clumps (RCs) was discovered in the high-latitude ($|b| \gtrsim 6\degr$) fields of the Milky Way bulge
\citep{mcw10,nat10}. The origin of this double RC, however, is currently under intense debate. It was initially interpreted as evidence
for an X-shaped structure that originated from the disk and bar instabilities \citep{mcw10,sai11,li12,weg13}.
In this picture, the two RCs separated by $\sim$0.5 mag is due to the difference in distance between the two arms of the X-shaped structure.
We have recently suggested \citep[hereafter \citetalias{lee15}]{lee15}, however, a drastically different interpretation based on the multiple stellar population phenomenon,
which is widely observed in globular clusters (GCs) including metal-rich bulge GCs Terzan~5, NGC~6388, and NGC~6441 \citep{lee99,cal07,yoo08,car09b,fer09,gra12}.
In this model, the brighter RC (bRC) is from helium enhanced second-generation stars (G2), which are intrinsically brighter,
while the faint RC (fRC) is originated from the first-generation stars (G1) with normal helium abundance.\footnotemark[1]

\footnotetext[1]{
As discussed in Paper I, we are not arguing against the bar dominated bulge in the low latitude ($|b| \lesssim 6 \degr$) fields.
The question is whether this pseudo bulge characteristic is extended even to the high latitude region as suggested by the X-shaped structure interpretation
of the double RC phenomenon. In our model, low-latitude fields are dominated by the most metal-rich bar population, while relatively metal-poor classical bulge
(CB) population becomes more and more important at higher latitudes. While the low latitude fields show cylindrical rotation,
stars in high latitude fields rotate more slowly \citep{zoc14}. Furthermore, \citet{sah12} showed that an initially non-rotating CB
could absorb a significant fraction of the angular momentum from the bar within a few Gyr, which suggests that the cylindrical rotation is not necessarily
an evidence against the coexistence of CB. \citet{zoc14} also pointed out that many early-type galaxies and the bulges originated from clumps \citep{elm08}
are fast rotators. Note also that a three-dimensional density map of \citet{weg13} is based on the X-shaped structure interpretation of the double RC.
Therefore, whether the high latitude field is also dominated by the bar population or not depends largely on the interpretation of the double RC.
}

Soon after our suggestion, two counterarguments were presented. Firstly, the longitude dependence of the double RC feature at $b \approx -8\degr$
was pointed out by \citet{gon15} as a potential problem of our model. Secondly, by employing the WISE mid-IR image,
\citet{nes16} report a direct detection of a faint X-shaped structure in the Milky Way bulge. However, each of them can be rebutted.
In a composite bulge, where a classical bulge is embedded in a tilted bar \citep[e.g.,][]{bab10,hil11,sah12,sah16,roj14,zoc14,erw15,sah15},
we have already shown in \citetalias{lee15} that our models can reproduce the longitude dependence well,
which is also discussed in Section~3 of this paper with more realistic treatments.
The second argument on the presence of a faint X-shaped structure is highly questioned as well, because when an ellipsoid is subtracted from a boxy structure,
as has been done by \citet{nes16}, an artificial X-shaped structure always remains (D. Han \& Y.-W. Lee 2017, in preparation; see also \citealt{lop16}).
Furthermore, if the X-shaped structure is solely responsible for the observed double RC, $\sim$45$\%$ of stars in the bulge are predicted
to be on the orbits that make this structure \citep{por15}. However, even if real, the stellar density in the claimed X-shaped structure of \citet{nes16}
appears to be way too low to be observed as a strong double RC at $l=0\degr$.
A crucial evidence against the X-shaped bulge was also presented by \citet{lop16} from the analysis of the main-sequence (MS) stars in the bulge fields.
Without critical distance information, therefore, the origin of
the double RC phenomenon is still an open question, and the multiple population scenario merits more detailed investigations.
The purpose of this paper is to extend the parameter space in our multiple population models to see the effects of metallicity, age, and helium abundance
on the double RC feature. We have also used these models to illustrate how the point raised by \citet{gon15} can be explained in our scenario.

\advance\textheight +24pt

\section{POPULATION MODELS}

Population models presented in this study are based on \citetalias{lee15} and the techniques previously developed by \citet{lee90,lee94} and \citet{joo13}.
As in \citetalias{lee15}, the Yonsei-Yale (Y$^2$) isochrones and horizontal-branch (HB) evolutionary tracks \citep{yi08,han09} were used to construct
the models for the normal and helium-enhanced subpopulations, with the assumption of the fixed $\alpha$-elements enhancement ([$\alpha/\rm Fe]=0.3$).
Parameters regarding the HB models, such as \citet{rei77} mass-loss coefficient ($\eta=0.40$) and
the mass dispersion on HB stars ($\sigma_M = 0.010 \rm M_\odot$), are also identical to those adopted in~\citetalias{lee15}.
Color-temperature transformation was performed by employing the color table of \citet{gre87} and the stellar model atmospheres of \citet{cas03}.
In order to more clearly see the effect of parameter changes on the synthetic RC, we present only single metallicity models in this paper.

\begin{figure}
\epsscale{1.0}
\plotone{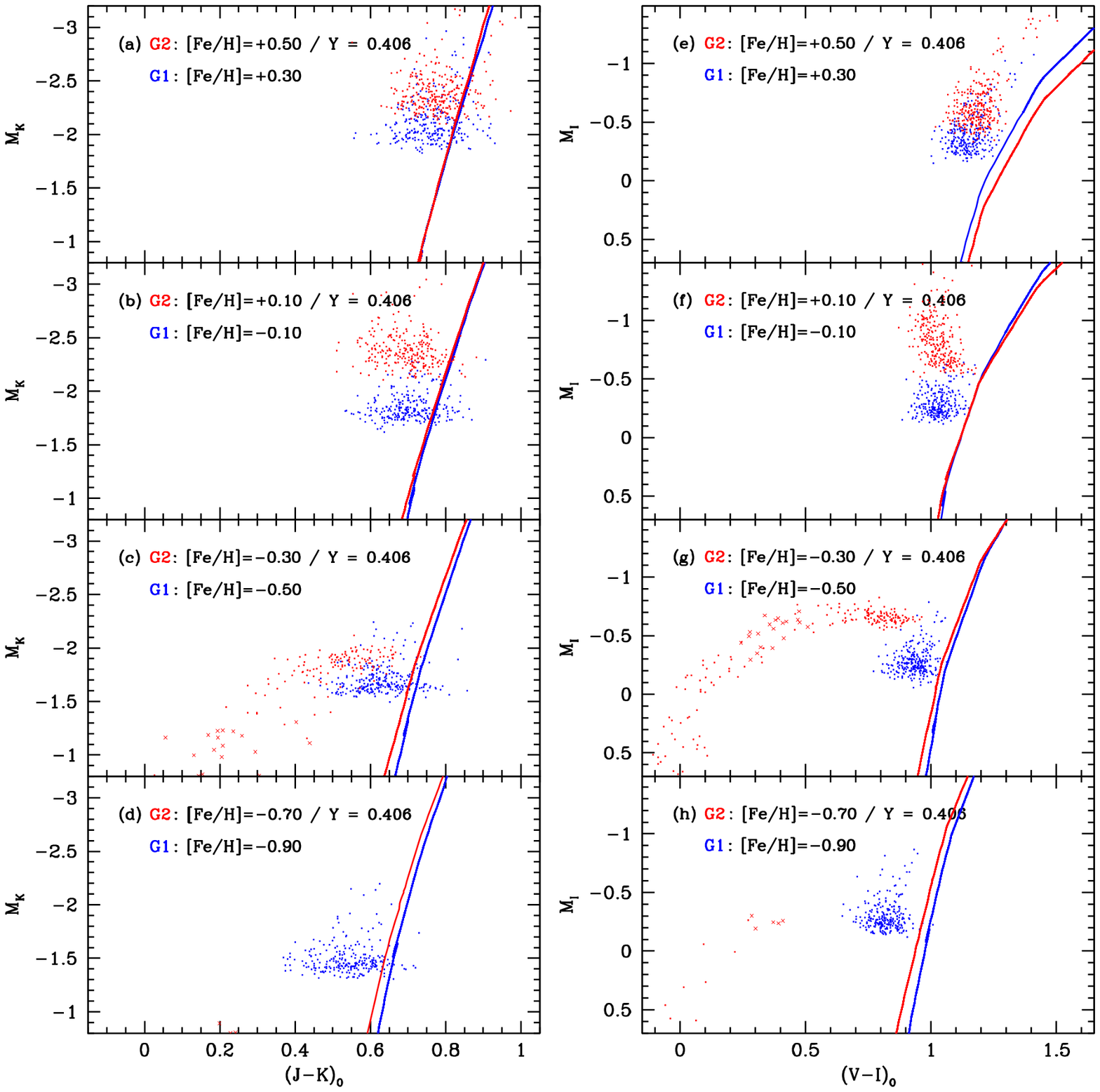}
\caption{
Our synthetic $(J-K,K)$ and $(V-I,I)$ CMDs for the two RCs at four different metallicity regimes. The fRC stars are produced by G1 with the standard helium-enrichment parameter
($\Delta Y/\Delta Z = 2.0$, $Y_p = 0.23$, $Y = Y_p + Z(\Delta Y/\Delta Z)$), and the bRC is from G2 with the substantially enhanced helium abundance, $Y=0.406$.
The left panels are almost identical to those presented in Figure~1 of \citetalias{lee15} \citep{lee15}, while the helium abundance of G2 was slightly revised from $Y=0.39$ to $0.406$.
Crosses are RR Lyrae variable stars from G2. Panels~(b) and (f) represent the ``$reference~models$", with [Fe/H]~=~$-$0.1, $t=12$~Gyr, and $Y = 0.276$ ($\Delta Y/\Delta Z = 2.0$)
for G1, and, [Fe/H] = +0.1, $t=10$~Gyr, and $Y=0.406$ for G2 (see the text).
\label{fig1}}
\end{figure}

\begin{figure}
\epsscale{1.0}
\plotone{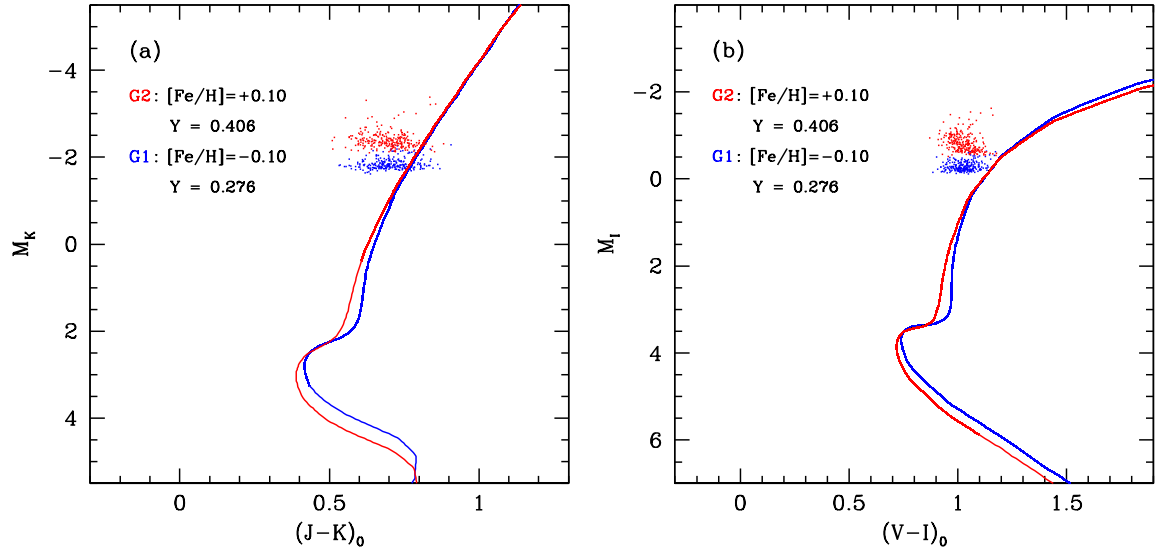}
\caption{
CMDs for our reference models, down to the MS luminosity level. These models are identical to those presented in panels~(b) and (f) of Figure~\ref{fig1} (see the text).
\label{fig2}}
\end{figure}

Figure~\ref{fig1} shows our synthetic color-magnitude diagrams (CMDs) for the two RCs at four different metallicity regimes,
which are almost identical to those presented in Figure~1 of \citetalias{lee15}, but here ($V-I, I$) CMDs are also added in the right panels.
We recall that, in our models, the fRC and bRC are produced by G1 and G2, respectively,
where G1 follows the standard helium-enrichment parameter (i.e., $\Delta Y/\Delta Z = 2.0$, $Y_p = 0.23$, $Y = Y_p + Z(\Delta Y/\Delta Z)$),
while G2 is substantially enhanced in helium abundance ($Y = 0.406$). Following \citetalias{lee15}, we have also adopted a 0.2 dex difference
in metallicity between G2 and G1, and assumed 12 and 10 Gyrs for the ages of G1 and G2 respectively, with the same population ratio for the two RCs
(\citealp{nat10}; \citetalias[][and references therein]{lee15}).
It is evident from these models that, in the metal-rich population like the bulge, highly helium-enhanced stars $(\rm Y \approx 0.40)$
are not placed on the very blue HB as in the metal-poor GCs, but are instead placed on the brighter RC.
We refer the reader to section~2 of \citetalias{lee15} for a detailed description of the RC features of our models
at four different metallicity regimes in Figure~\ref{fig1}. Here we note from the ($V-I, I$) CMDs that the variation
of the overall RC features on metallicity is similar to that in ($J-K, K$) CMDs.
Panel~(f) in ($V-I, I$) CMD can also naturally reproduce the observed double RC feature, i.e., $\sim$0.5 mag difference
with almost negligible color difference between the two RCs, as is the case in panel (b) in ($J-K, K$) CMD.
Hereafter, we refer to these models in panels~(b) and (f) constructed at $<$[Fe/H]$>$ = 0.0 as ``$reference~models$",
in order to distinguish them from the models with different input parameters presented below.
The CMDs for these reference models, down to the MS luminosity level, are further presented in Figure~\ref{fig2}.
This figure confirms that, in the position and width of the lower red-giant-branch (RGB) and main-sequence turn-off (MSTO), our models are
not inconsistent with the observed CMDs of bulge population by \citet{cla08,cla11} and \citet{bro10}.

While helium abundance was fixed for G2 ($Y = 0.406$) in the models in Figure~\ref{fig1},
a more realistic case, in terms of the chemical evolution, would be that the helium content varies with metal abundance
following a certain value of $\Delta Y/\Delta Z$. Models in Figure~\ref{fig3} were constructed
with this hypothesis, adopting $\Delta Y/\Delta Z = 6.0$ for G2. The models for G1 are identical to those in Figure~\ref{fig1}
following $\Delta Y/\Delta Z = 2.0$.
Note that the reference models in panels~(b) and (f) of Figure~\ref{fig3} are still identical to those in Figure~\ref{fig1}.
As the metallicity decreases from panels~(b) and (f) to (c) and (g), magnitude differences between the two RCs decrease drastically,
and in the bottom panels, the two RCs are almost merged into one.
This is due to the strong decrease in helium abundance for G2 following $\Delta Y/\Delta Z = 6$.
Compared to the models in Figure~\ref{fig1}, where the helium abundance of G2 is fixed, the bRC stars
in lower panels in Figure~\ref{fig3} are confined in the RC region rather than shifted to the bluer HB regime.
Consequently, in both cases (i.e., Figures~\ref{fig1} and \ref{fig3}), no double RC is predicted by our models
in the metal-poor regime, although the HB morphologies produced by G2 are significantly different in the two cases (see also \citetalias{lee15}).
According to the chemical evolution models \citep[e.g.,][]{bek06,rom07}, helium abundance of G2 can not increase
beyond $Y_{\rm max} \approx 0.41$; therefore, we adopted this value for G2 in the most metal-rich models in panels~(a) and (e).
Accordingly, the magnitude differences between the two RCs decrease again in these panels.

In order to see the individual effects of age, helium abundance, and metallicity on the double RC feature,
in Figures \ref{fig4} to \ref{fig7}, we have constructed the RC models under different assumptions on these parameters for G1 and G2.
In these figures, luminosity functions are also presented on the right-side of each CMD, which were derived from $\sim$10,000 stars
for G1 and G2 respectively to minimize a stochastic effect.
In Figure~\ref{fig4}, we first investigate the effect of age variation on G2 from the reference models.
From top to bottom panels, the age for G2 is decreased from 12 to 6~Gyrs, while the age for G1 is fixed (12 Gyr),
so that the age difference between G1 and G2 ($\Delta t_{\rm G1-G2}$) increases from 0 to 6~Gyrs.
The helium abundance and metallicity are also fixed for both G1 and G2 as in the reference models.
In the top panels, where the age for G2 is increased by 2~Gyr from the reference models in panels~(b) and (f), it is evident that
some significant portion of bRC stars are shifted to the blue HB, making the bluer mean color for the bRC stars
compared to the stars in fRC. As such, this could be considered as a reasonable upper limit for the age of G2,
at the adopted metallicity and helium abundance.
When the age of G2 is decreased by 2$-$4~Gyrs from the reference models; however, as shown in the four bottom panels,
the overall bRC feature is not remarkably changed, implying an insensitivity of the bRC morphology in this age range.
Compared to the reference models in panels~(b) and (f), the stars in bRC in the bottom panels are $\sim$0.2 mag brighter and $\sim$0.1 mag redder,
while the age of G2 is decreased by 4~Gyrs, which would give a lower limit for the age of G2.

In Figure~\ref{fig5}, we examine the case where the age of G1 also varies from 14 to 6 Gyr, from the top to bottom panels,
while the ages for G2 are identical to those in Figure~\ref{fig4}. Again, panels~(b) and (f) present the reference models.
In the top panels, where the age of G1 is increased by 2~Gyr from the reference models, no obvious changes are noticed in the fRC morphology,
while the mean luminosity of fRC stars is slightly decreased ($\lesssim$ 0.05 mag) in both $K$ and $I$-bands.
Similarly, when the age of G1 is decreased to be 2-6 Gyr younger than that of the reference model, in the four bottom panels,
no apparent changes in the color of the fRC stars are detected, while their luminosity is slightly increased.
Consequently, in all panels, the fRC stars are confined to the RC region, even if they are as old as 14 Gyr,
illustrating that the color of the fRC is almost completely insensitive to the age variation in this high-metallicity regime.
The luminosity of the fRC stars is also not seriously affected by the age changes, although they become slightly brighter as age decreases.
From Figures~\ref{fig4} and~\ref{fig5}, we can therefore conclude that the possible ranges for age are
12 $\gtrsim t \gtrsim$ 6 Gyr for G2 and t $\gtrsim$ 6 Gyr for G1.
It should be noted that these age estimations are affected by the current uncertainty in the stellar mass-loss on the RGB.
For example, if we increase the \citet{rei77} mass-loss coefficient, $\eta$, from 0.40 to 0.50,
these ages are decreased by $\sim1.0-1.5$~Gyr.

The effect of helium abundance is illustrated in Figure~\ref{fig6}. In these models, helium mass fraction for G2 is varied from 0.41 to 0.35,
while all the other parameters are fixed as in the reference models.
The difference in helium abundance between the two RCs ($\Delta Y_{\rm G2-G1}$) is then decreased from 0.13 to 0.07.
The most prominent feature in this figure is that the magnitude difference between the two RCs decreases substantially as $\Delta Y_{\rm G2-G1}$ decreases,
from top to bottom panels. It is therefore clear from the luminosity functions that the observed separation in magnitude is reproduced in the top four panels,
but not in the bottom four panels. The color of the bRC stars, on the other hand, is less affected by the variation
in helium abundance, though they are somewhat redder than the fRC stars in the bottom four panels.
This is because G2 in our models are more metal-rich than G1, and thus the helium effect is mostly canceled out
by the metallicity effect in this metal-rich regime (see section~1 in \citetalias{lee15}).
Our simulations also suggest that, if the helium abundance for G2 is allowed to increase beyond the upper limit ($Y_{\rm max} = 0.41$)
adopted in our models, the colors of the bRC can be bluer than those in the top panels.
Consequently, the strong dependence of the double RC feature on the helium abundance of G2 clearly suggests that
the $\Delta Y_{\rm G2-G1}$ would be within the range of $\simeq$ 0.10 $-$ 0.13, in order to reproduce the observation in our scenario.

Figure~\ref{fig7} explores the cases where the metallicity difference between G2 and G1 ($\Delta \rm [Fe/H]_{G2-G1}$) is different
from our standard assumption, $\Delta \rm [Fe/H]_{G2-G1}$ = 0.2 dex. From top to bottom panels, metallicity for G2 is increased from [Fe/H] = $-$0.1 to +0.5,
so that $\Delta \rm [Fe/H]_{G2-G1}$ increases from 0.0 to 0.6 dex. The other parameters are fixed as in the reference models presented in panels~(b) and (f).
The most notable feature from these models is that the color of bRC considerably shifts from blue to red in both $J-K$ and $V-I$ colors
as metallicity for G2 increases. It is therefore clear that the difference in color between the two RCs changes substantially with $\Delta \rm [Fe/H]_{G2-G1}$.
Some decrease in luminosity of bRC is also shown in panels~(a), (g), and (h),
but the magnitude difference is not seriously affected by the metallicity variation in other panels.
From these, we conclude that the $\Delta \rm [Fe/H]_{G2-G1}$ would be no more than $\sim$0.3 dex to reproduce the observed double RC with negligible color difference.
Note that this is consistent with the metallicity difference ($\Delta \rm [Fe/H] = 0.12-0.23$~dex) between the two RCs
inferred from the spectroscopic observations \citep{dep11,nes12,utt12}.

It is also interesting to see that the double RC feature in panels~(c) and (g) of Figure~\ref{fig7} are qualitatively similar to that of Terzan~5,
a metal-rich GC in the bulge \citep{fer09,dan10,ori11}.
This suggests that a larger difference in metallicity between the two RC populations
is mainly responsible for the color difference between the two RCs in this GC (see also \citetalias{lee15}),
while a larger age difference \citep{fer09} would also help to increase the color difference.
Recently, \citet{fer16} have shown that the age of G2 in Terzan~5 is only $\sim$4.5~Gyr
from the detection of brighter MSTO stars, in addition to the faint majority MSTO stars of G1 with an old age ($\sim$12~Gyr).
In their comparison with stellar models, however, \citet{fer16} adopted $\Delta \rm [M/H]$ = 0.51 dex between G2 and G1,
which is too large considering the observed difference (0.3~dex) in $[\alpha \rm /Fe]$ between G1 and G2.
When this difference in $[\alpha \rm /Fe]$ is taken into account, $\Delta \rm [M/H]$ must be reduced to $\sim$0.29 dex.
Although G2 in Terzan~5 is much younger, the observed differences in age and metallicity are then not sufficient enough to reproduce the observed double RC
as illustrated by our models in Figure~\ref{fig8}. We still need a large difference in helium abundance $(\Delta \rm Y = 0.07)$.
Therefore, Terzan~5 is qualitatively similar to our multiple population model for the double RC in the bulge.
Since G2 in Terzan~5 is much younger, they are more affected (diluted) by SNe Ia, which would explain lower helium abundance (0.33 vs. 0.39) and
$[\alpha \rm /Fe]$ (0.0 vs. 0.2) compared to bulge field stars in our scenario.
In any case, Terzan~5 shows directly that multiple populations can produce a double RC.

In the metal-poor populations like the Milky Way GCs, the HB morphology is well known to be crucially influenced by all the three parameters (age, helium abundance,
and metallicity) explored above \citep[e.g.,][]{lee90,lee94}.
Our model simulations indicate, however, that in the metal-rich regime ([Fe/H] $\approx 0.0$), the HB stars are mostly confined in the RC region,
where the effects of these parameters are either qualitatively or quantitatively different from the metal-poor cases.
The luminosity of RC stars is mainly affected by helium abundance, while the RC color is primarily affected by metallicity.
The effect of age is relatively small within the age range of $12 \gtrsim t \gtrsim 6$.
From our models, we can suggest the possible ranges of parameters required to reproduce the observed double RC feature:
$\Delta t_{\rm G1-G2}$ $\approx 0 - 6$ Gyr, $\Delta Y_{\rm G2-G1}$ $\approx 0.10 - 0.13$,
and $\Delta \rm [Fe/H]_{G2-G1}$ $\approx 0.0 - 0.3$ dex.
It should be noted, however, that these three parameters are not independent from each other but degenerated.
For example, Figure~\ref{fig9} illustrates how the different combinations of parameters could reproduce the observed double RC feature.
From the top to bottom panels, we increased the metallicity of G2 from [Fe/H] = $-$0.1 to +0.2,
so that the metallicity difference between G2 and G1 increases from $\Delta \rm [Fe/H]_{G2-G1}$ = 0.0 to 0.3, while the models for G1 are
fixed as the reference models.
We have then adjusted helium abundance and age for G2 within the parameter space suggested above,
until the models in each panel similarly reproduce the double RC feature of the reference models
presented in panels~(c) and (g). Note that the helium abundances of G2 are significantly ($\Delta Y_{\rm G2-G1} \geq 0.12$) enhanced in all panels.
Our models indicate, therefore, that the observed double RC feature can be naturally generated in a relatively large parameter space
without the need for fine-tuning, once $\Delta Y_{\rm G2-G1}$ is assumed to be greater than $\sim$0.10.

\section{LONGITUDE DEPENDENCE AT $\MakeLowercase b \approx -8\degr$}

\advance\textheight -90pt

\citet{gon15} pointed out that our multiple population models for the double RC can not explain the longitude dependence of the RC luminosity function
at high Galactic latitude, $b \approx -8\degr$. As shown in their Figures~1 and 2, the population in bRC becomes more prominent
towards the positive longitudes, while the fRC population becomes more dominant in the negative longitudes.
If this high-latitude field is dominated by the classical bulge population with negligible contribution from the bar component, as argued by \citet{gon15},
it would be challenging for our models to reproduce the longitude dependence. However, in the case where the bar component is still not negligible
over $b \approx -8\degr$, our models can reproduce such lateral variations by a combination of the classical bulge and bar populations,
as shown in Figure~4 of \citetalias{lee15}. In order to investigate this effect in a composite bulge in more detail, we further conduct Monte-Carlo simulations
for the three dimensional density distributions of the classical bulge and the bar components.
For the classical bulge component, we adopt the spatial density distribution of bulge RR Lyrae stars, ${\rho} \propto r_{z}^{-2.9}$, from \citet{pie15},
and assume a simple spherical shape \citep{dek13}.\footnotemark[2] Following \citet{gar10} and \citet{rom11},
the bar component is modeled by a simple Ferrers ellipsoid \citep{fer77} with a density function of ${\rho} \propto (1-m^2)^n$,
where $m^2=x^2/a^2+y^2/b^2+z^2/c^2$ and $n=1$ when $m \leq 1$ ($\rho$ becomes 0 when $m > 1$).
We assume the COBE/DIRBE bar with the axis ratio of $a:b:c=10:4:3$ \citep[][]{wei94, fre98, ger02}, and the position angle of the major axis
with respect to the Galactic Center-Sun line of 30$\degr$ \citep[][]{lop05, rom11}.

\footnotetext[2]{
\citet{kun16} found that RR Lyrae stars ($\rm [Fe/H] \approx -1$) rotate more slowly than the most metal-rich RC and RGB stars,
and suggest that their kinematics and spatial distribution are consistent with a CB. \citet{zoc16} confirm that this trend is extended
to more metal-rich ($\rm [Fe/H] < -0.1$) population. They found that metal-poor ($\rm [Fe/H] < -0.1$) RC and RGB stars rotate slower than metal-rich stars ($\rm [Fe/H] > +0.1$).
At high latitude ($|b| \gtrsim 4.5\degr$) fields, they also show that metal-poor population has a larger velocity dispersion than metal-rich stars.
Furthermore, they found that the metal-poor population is more centrally concentrated with a more axisymmetric spatial distribution,
while metal-rich population shows a boxy distribution. While \citet{zoc16} interpreted these observations differently, all of these characteristics
are consistent with the two component bulge (CB+bar) assumed in this paper. Furthermore, as discussed in Paper I (Section 3),
the observed kinematics of bulge RC and RGB stars are not inconsistent with our composite bulge scenario.
}

\advance\textheight +90pt

Figure~\ref{fig10} shows density maps of our simulations for the two component bulge model, where the classical bulge, the bar, and the sum of these two components
are displayed from the top to bottom panels. Since we adopted a simple Ferrers bar, our composite model (the bottom right panel)
may appear not as boxy as the observed bulge \citep[e.g.,][]{lop05}.
Note, however, that our models are not for the detailed comparison with the observed density map,
but to illustrate the variation of the population ratio between the classical bulge and the bar components with galactic longitude and latitude,
and the corresponding changes in the RC luminosity function.
As illustrated by our simulations, if the classical bulge component has a spheroidal shape, the stellar density monotonically decreases
with increasing distance from the Galactic center, while the density of the bar mostly depend on the distance above the Galactic plane.
Therefore, at fixed latitude, the classical bulge to bar population ratio would reach the maximum at minor axis ($l = 0\degr$) and decrease as longitude ($|l|$) increases.

In Figure~\ref{fig11}, we present this effect on the RC luminosity function at $b\simeq -5$ and $-8\degr$ based on the stellar density distributions in Figure~\ref{fig10}.
For the direct comparison with Figure~2 of \citet{gon15}, we chose three fields at Galactic latitudes of $l=+2$, $0$, and $-2\degr$.
For the luminosity function of the classical bulge component, we adopt $\Delta K \simeq 0.6$~mag between the two RCs, and a Gaussian noise is added
to reflect the differential reddening \citep{gon12} and other minor effects at chosen fields for both classical bulge and bar components
($\sigma_{K}=0.05$ for $b=-8\degr$ and 0.15 for $-5\degr$).
The large dispersions in RC luminosity functions are mainly due to the magnitude differences that come purely from distance differences of each star from the Sun.
In our model, the RC luminosity function would be the sum of the classical bulge component with double RC and the bar component with a single RC.
The double RC in the lower middle panel ($b=-8\degr$) is due to the majority classical bulge population, while the longitude variations
of the RC luminosity function, in the left and right panels, are mostly produced by the distance effect of the minority population of the tilted bar,
where the bar component is placed at the near (far) side of the bulge towards positive (negative) longitude.
This effect of the bar is more evident in the fields at $b=-5\degr$ where the contribution of the bar population is stronger.
The magnitude difference due to the distance effect is about $\Delta K \approx 0.09$~mag between the near and far sides of the bar component.
This leads to the skewed distribution of RC luminosity function as the longitude $|l|$ of the field increases (see top panels of Figure~\ref{fig11}).
In addition, when the tilted bar is projected on the sky, the near-side of the bar at positive longitude would appear larger than the far-side component
at negative longitude (see the middle right panel in Figure~\ref{fig10}).
Therefore, at the same latitude, towards the positive longitude, we would observe the bulge field at a lower distance above the plane,
where the stellar density is expected to be higher, and consequently, the bar contribution would be more prominent in the positive longitude
compared to the negative longitude. This effect is also included in Figure~\ref{fig11}.
Note also that the luminosity function of the bar component at $b=-8\degr$ is shifted toward the brighter side, which is because the line of sight
passes through the near boundary of the Ferrers bar. It is encouraging to see that the observed RC luminosity functions
presented in Figure~2 of \citet{gon15} are better explained by our simulations than their X-shaped bulge model, especially at $l=-2\degr$.

There are at least two possible alternative effects that can also help to explain the longitude dependence at $|b| \approx 8\degr$ fields.
The first effect comes from the recent analysis of RR Lyrae variable stars in the Galactic bulge by \citet[][see their Fig.~6]{pie15}, which shows
that even the classical bulge component is elongated in the inner region with the shape of a triaxial ellipsoid following the bar structure,
while the outer region shows more spherical distribution. The RR Lyrae luminosity function in Figure~6 of \citet{pie15} shows some hint
of asymmetric distribution in the luminosity function with longitude dependence.

The other effect is from the possible metallicity variation with longitude at $b \approx -8\degr$.
According to the metallicity map of \citet{gon13}, the mean metallicity of the bulge field at $b \approx -8\degr$
decreases from $<$[Fe/H]$>$ $\approx -$0.2 to $-$0.5 as the longitude varies from $l=0\degr$ to $-4\degr$.
Following this metallicity map, in Figures~\ref{fig12} and \ref{fig13}, we have investigated how our models are affected by this metallicity variation.
In this metallicity scale of \citet{gon13}, the double RC is best reproduced when $\Delta$Y/$\Delta$Z is adopted to be 8.0 for G2 and
ages for G1 and G2 are assumed to be 11 and 9 Gyrs, respectively. From the top to bottom panels,
we have then decreased the metallicity by 0.1 dex with the other parameters held fixed,
which would mimic the metallicity variation from $l=0\degr$ to $\sim$$-4\degr$.
Similarly to the metallicity dependence presented in Figure~\ref{fig3}, the luminosity of bRC stars decreases rapidly with decreasing metallicity,
and, with the metallicity change of only $\Delta$[Fe/H] = 0.2 dex, the RC split disappears as shown in panels~(c) and (g).
As discussed above, this drastic change of the bRC luminosity is mostly due to the large variation of helium abundance according to the high value of
$\Delta$Y/$\Delta$Z parameter.
Therefore, if the metallicity variation is confirmed, this effect would also help to reproduce the longitude dependence of the RC luminosity function
on the negative longitude side. On the positive longitude side, the stronger peak at the bRC would be explained by the projection effect discussed above.
The point raised out by \citet{gon15} is therefore not necessarily a potential problem in our models,
since our multiple population models can equally reproduce the observed variation of luminosity function with longitude at $b \approx -8\degr$.

\section{DISCUSSION}

\begin{figure}
\epsscale{1.0}
\plotone{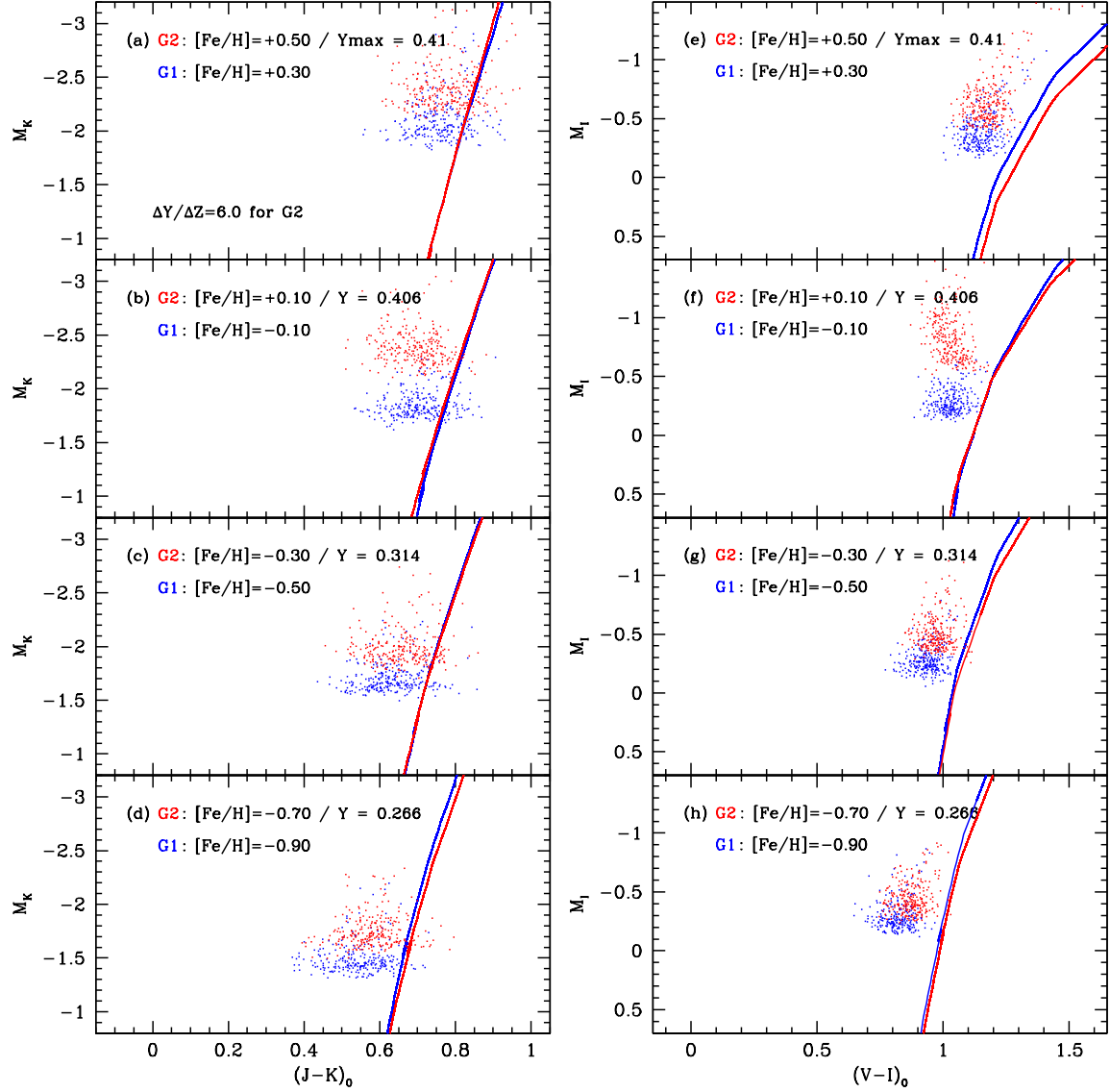}
\caption{
Same as Figure~\ref{fig1}, but the helium abundance for G2 varies with metallicity following $\Delta Y/\Delta Z = 6.0$.
Note that the magnitude difference between the two RCs decreases as $<$[Fe/H]$>$ decreases.
For the most metal-rich models in panels~(a) and (e), we adopt $Y_{\rm max}=0.41$ as the maximum value for G2 (see the text).
Panels~(b) and (f) are identical to the reference models in Figure~\ref{fig1}.
\label{fig3}}
\end{figure}

\begin{figure}
\epsscale{1.0}
\plotone{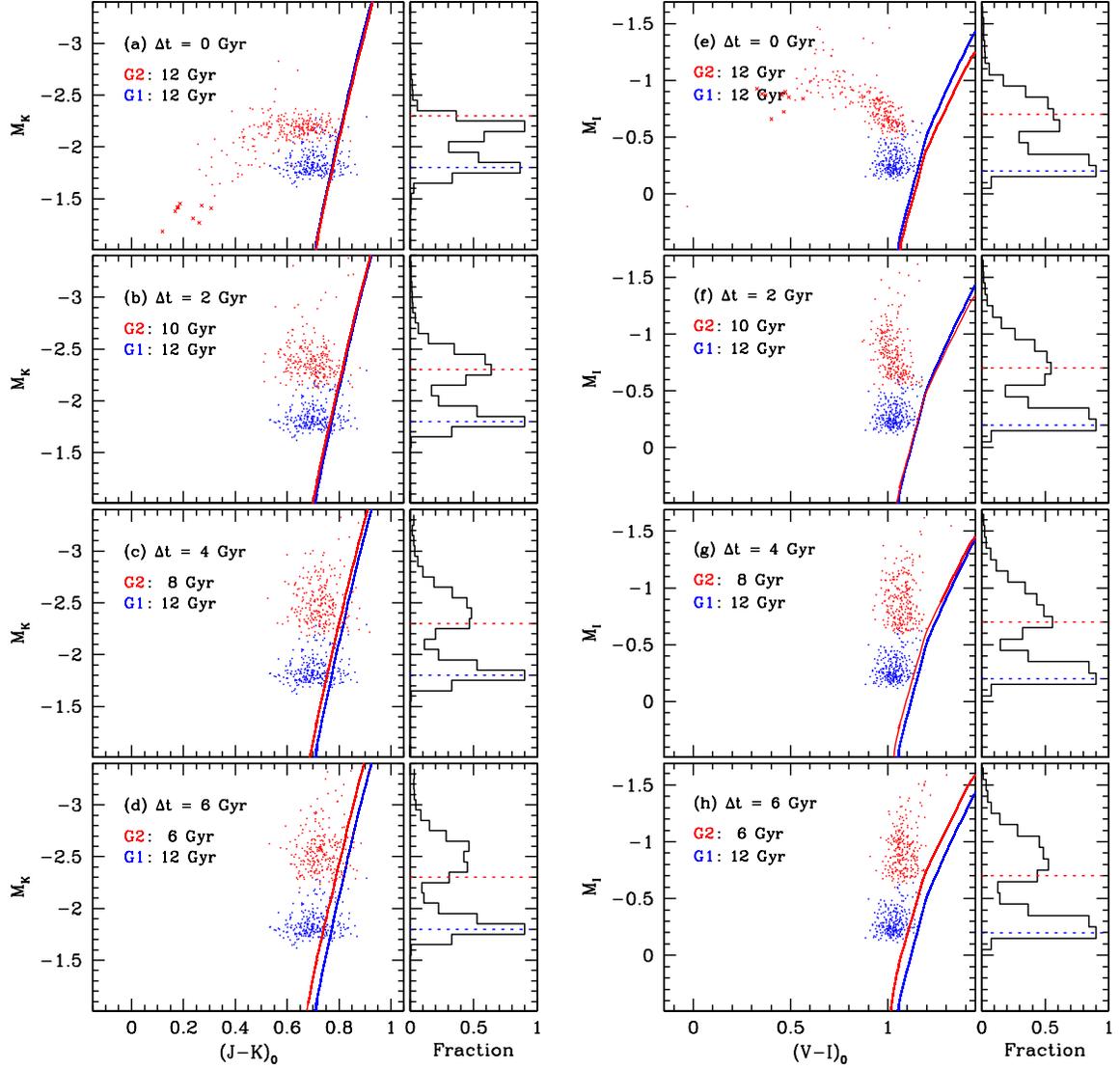}
\caption{
Effect of age variation on G2 in our synthetic CMDs and luminosity functions for the two RCs. From top to bottom panels,
the age for G2 is decreased from 12 to 6 Gyrs, while all the other parameters (for both G1 and G2) are held fixed as in the reference models in panels~(b) and (f).
The horizontal dotted lines represent two peaks of the RC luminosity function for the reference models.
\label{fig4}}
\end{figure}

\begin{figure}
\epsscale{1.0}
\plotone{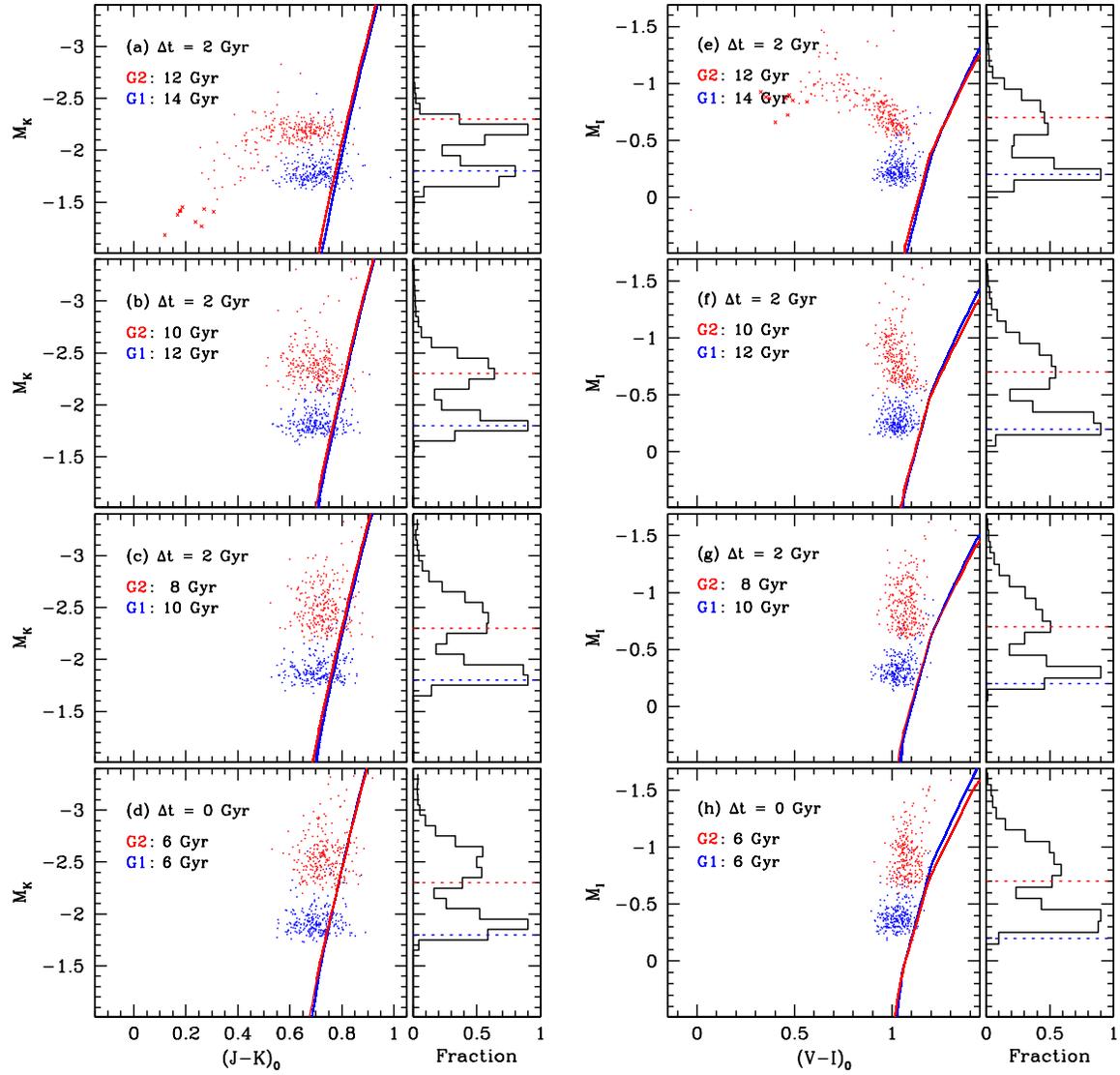}
\caption{
Same as Figure~\ref{fig4}, but for the effect of age variation on G1. From top to bottom panels, the age for G1 is also decreased from 14 to 6 Gyr,
while the adopted ages for G2 are identical to those in Figure~\ref{fig4}. The metallicites and helium abundances for both G1 and G2 are
fixed as in the reference models in panels~(b) and (f).
\label{fig5}}
\end{figure}

\begin{figure}
\epsscale{1.0}
\plotone{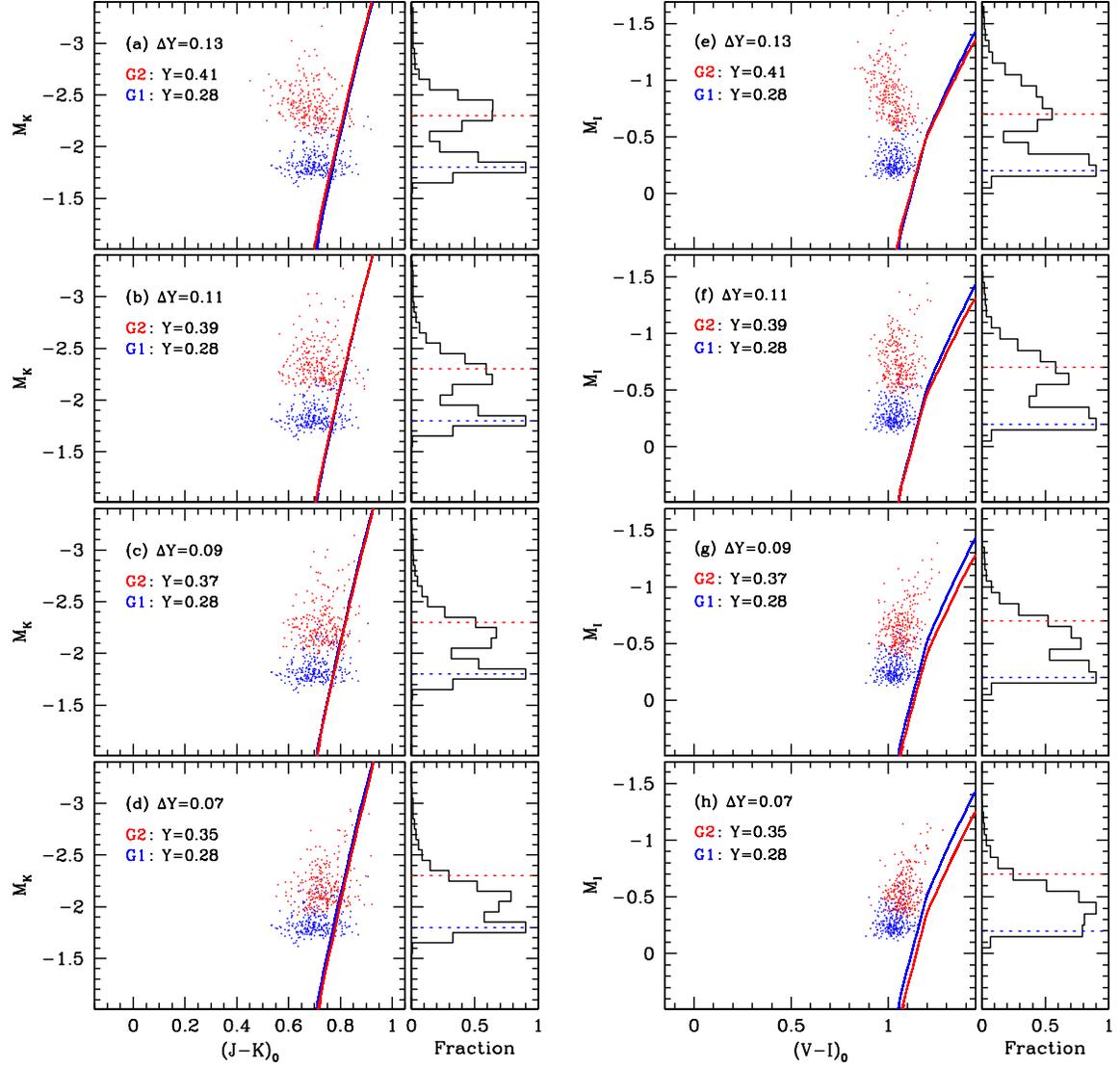}
\caption{
Similar to Figure~\ref{fig4}, but for the effect of helium abundance variation on G2. The helium abundance for G2 is changed from 0.41 to 0.35,
while the other parameters are fixed as in the reference models. Note that $\Delta \rm Y \gtrsim 0.10$ is required to reproduce the observed double RC feature.
Panels~(a) and (e) are almost identical to the reference models.
\label{fig6}}
\end{figure}

\begin{figure}
\epsscale{1.0}
\plotone{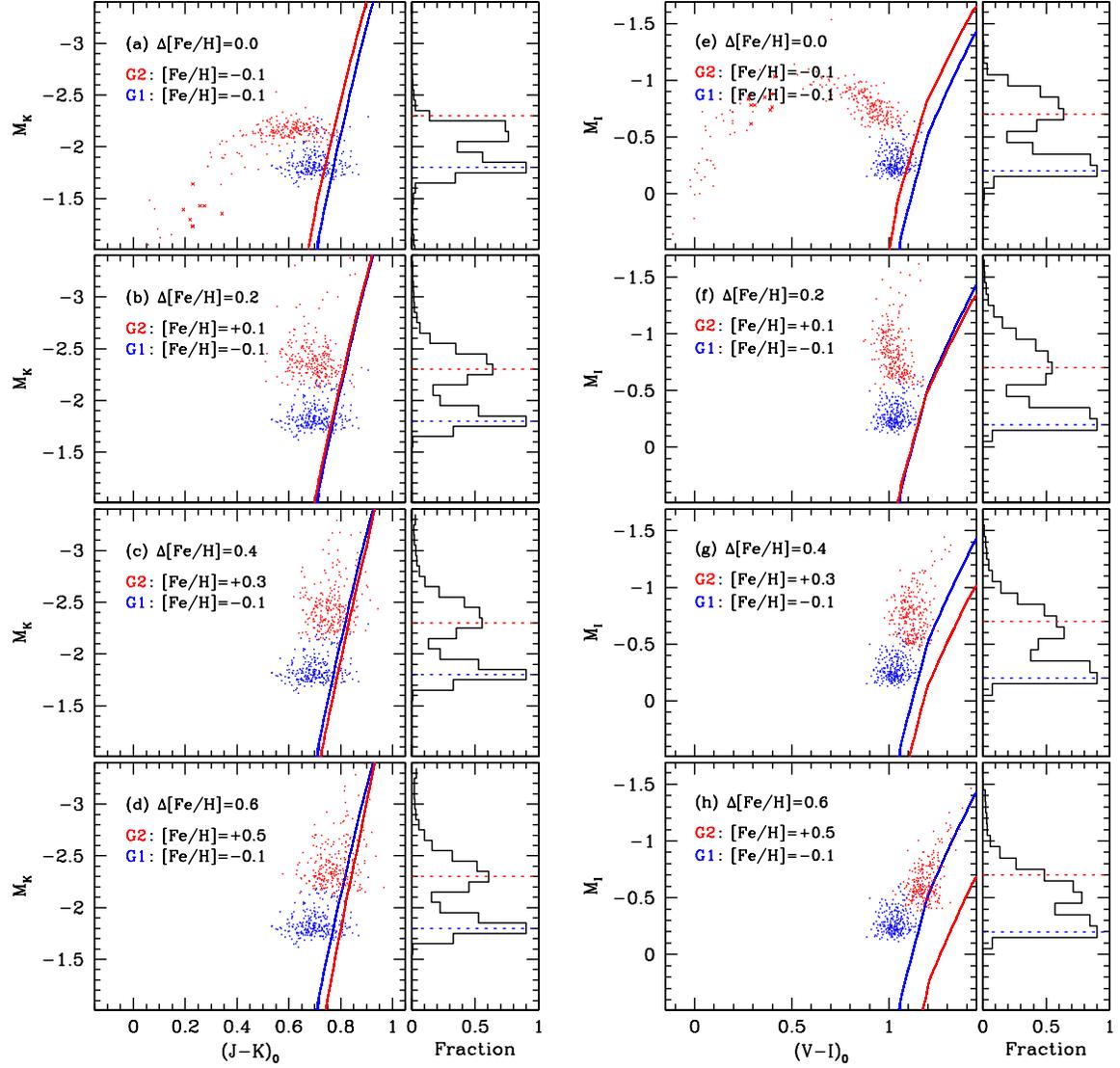}
\caption{
Similar to Figure~\ref{fig4}, but for the effect of $\Delta$[Fe/H] variation between G2 and G1. From top to bottom panels,
[Fe/H]$_{\rm G2}$ is increased from $-$0.1 to +0.5, while [Fe/H]$_{\rm G1}$ is fixed at $-$0.1, and thus,
$\Delta \rm [Fe/H]_{G2-G1}$ increases from 0.0 to 0.6 dex. All the other parameters are fixed as in the reference models in panels~(b) and (f).
\label{fig7}}
\end{figure}

In the multiple population paradigm, we have shown that the magnitude difference between the two RCs is mostly affected by the difference
in helium abundance between G2 and G1, and the observed double RC feature can be reproduced if $\Delta Y_{\rm G2-G1} \approx 0.10-0.13$ at around solar metallicity.
As illustrated in Figure~\ref{fig3}, this is obtained if the helium abundance for G2 increases with metallicity following a helium enrichment parameter $\Delta Y/\Delta Z = 6.0$,
while G1 follows a standard value $\Delta Y/\Delta Z = 2.0$. Chemical evolution models (J. Kim \& Y.-W. Lee 2017, in preparation) confirm that
this trend for G2 is predicted if the gas that formed G2 was locally enriched within the protogalactic subsystems by the winds of massive stars and
the winds and ejecta from low and intermediate mass asymptotic-giant-branch stars. Disruption of these ``$building~blocks$"
in a hierarchical merging paradigm would have provided stars (G1 and G2) to form a classical bulge component in the early stage of the Milky Way formation.
One crucial condition required to obtain this result is to assume that most of the supernova ejecta could escape from these relatively less massive systems
without expelling the pre-enriched gas inside \citep[see, e.g.,][]{rom10,ten15}. Because of the strong metallicity dependence of helium yield
from the winds of massive stars \citep{mae92,mey08}, helium abundance of G2 in these models would then increase rapidly with metallicity.

In our models, the helium abundance of G2, following $\Delta Y/\Delta Z = 6.0$, would be relatively low $\rm (Y = 0.24 - 0.25)$ at metal-poor regime
(i.e., halo GCs), while it is very high $\rm (Y \approx 0.40)$ at solar metallicity where the double RC is observed (see Fig. 11 of \citealt{lee16}).
Therefore, G2 in our models would correspond to ``$Na-intermediate$" population defined by \citet{car09a}.
Very Na-rich and O-poor stars (``$Extreme$" population) are defined as G3 (3rd and later generations) in our models, which, if any,
are more likely to be trapped in bulge GCs as in the metal-poor halo GCs \citep{der08}, and thus we do not expect many of them in the bulge fields.
Nevertheless, at around solar metallicity, some significant spread in [Na/Fe] $(\Delta \rm [Na/Fe] \approx 0.4 - 0.5 ~dex)$
is observed in the bulge (see Figs. 15 \& 16 of \citealt{joh14}; Fig. 7 of \citealt{lec07}), which is clearly larger than that of the disk
$(\Delta \rm [Na/Fe] \approx 0.1 ~dex)$ and halo $(\Delta \rm [Na/Fe] \approx 0.25 ~dex)$ field populations (see Fig. 11 of \citealt{vil17}),
suggesting that the bulge contains more stars from G2 and later generations.
\citet{lec07} even reports some indication of Na-O anticorrelation, although this has not been confirmed by \citet{joh14}.

\begin{figure}
\epsscale{1.0}
\plotone{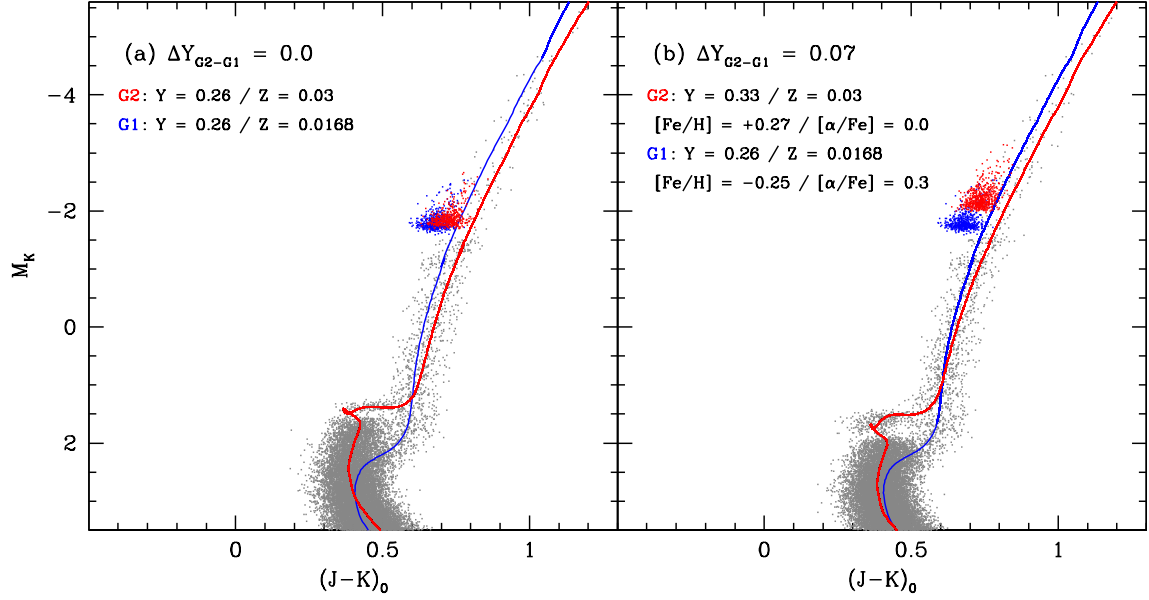}
\caption{
Our synthetic CMDs for the metal-rich GC Terzan~5 to illustrate the effect of helium abundance on G2. Our models clearly show that
a large enhancement in helium abundance $(\Delta \rm Y = 0.07)$ is needed for G2 to reproduce the observed double RC.
Following \citet{fer16}, we adopt 12 and 4.5 Gyrs for the ages of G1 and G2, respectively. Iron and $\alpha$-elements abundances are from \citet{ori11}.
Grey dots are synthetic CMDs for MS and RGB stars with photometric errors.
\label{fig8}}
\end{figure}

\begin{figure}
\epsscale{1.0}
\plotone{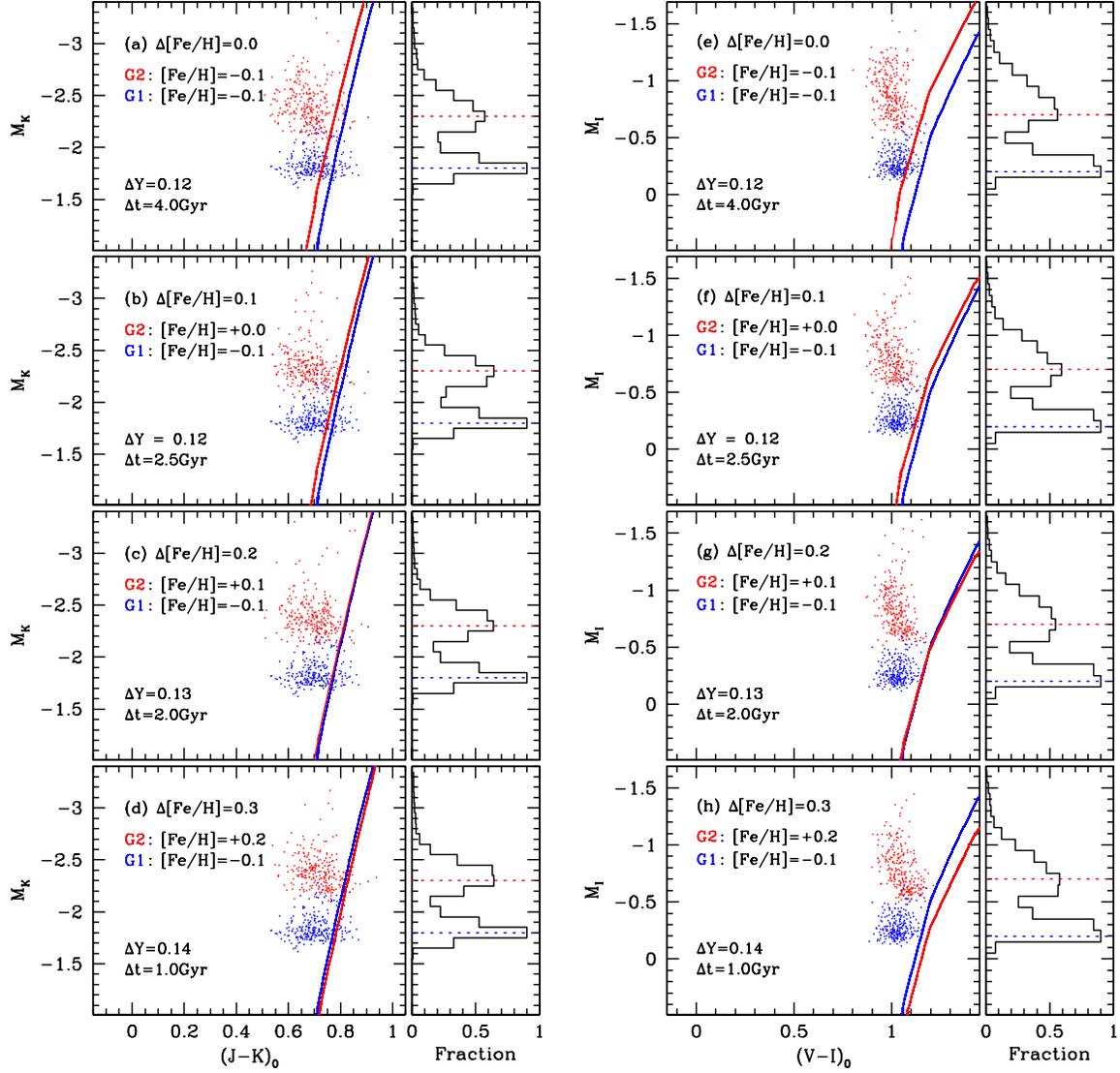}
\caption{
Similar to Figure~\ref{fig4}, but here the models are constructed with different combinations of parameters to reproduce the double RC feature.
While the model parameters for G1 are fixed as in the reference models, [Fe/H]$_{\rm G2}$ is increased from $-$0.1 to +0.2, from top to bottom panels,
so that $\Delta \rm [Fe/H]_{G2-G1}$ varies from 0.0 to 0.3. The helium abundances and ages of G2 are then adjusted until we get reasonable agreements
with the reference models in panels~(c) and (g). The adopted values are denoted in the bottom left corner of each panel.
Note that, in all panels, $Y_{\rm G2}$ is significantly enhanced ($\Delta Y_{\rm G2-G1} \geq 0.12$).
\label{fig9}}
\end{figure}

\begin{figure*}
\includegraphics[angle=-90,scale=0.70]{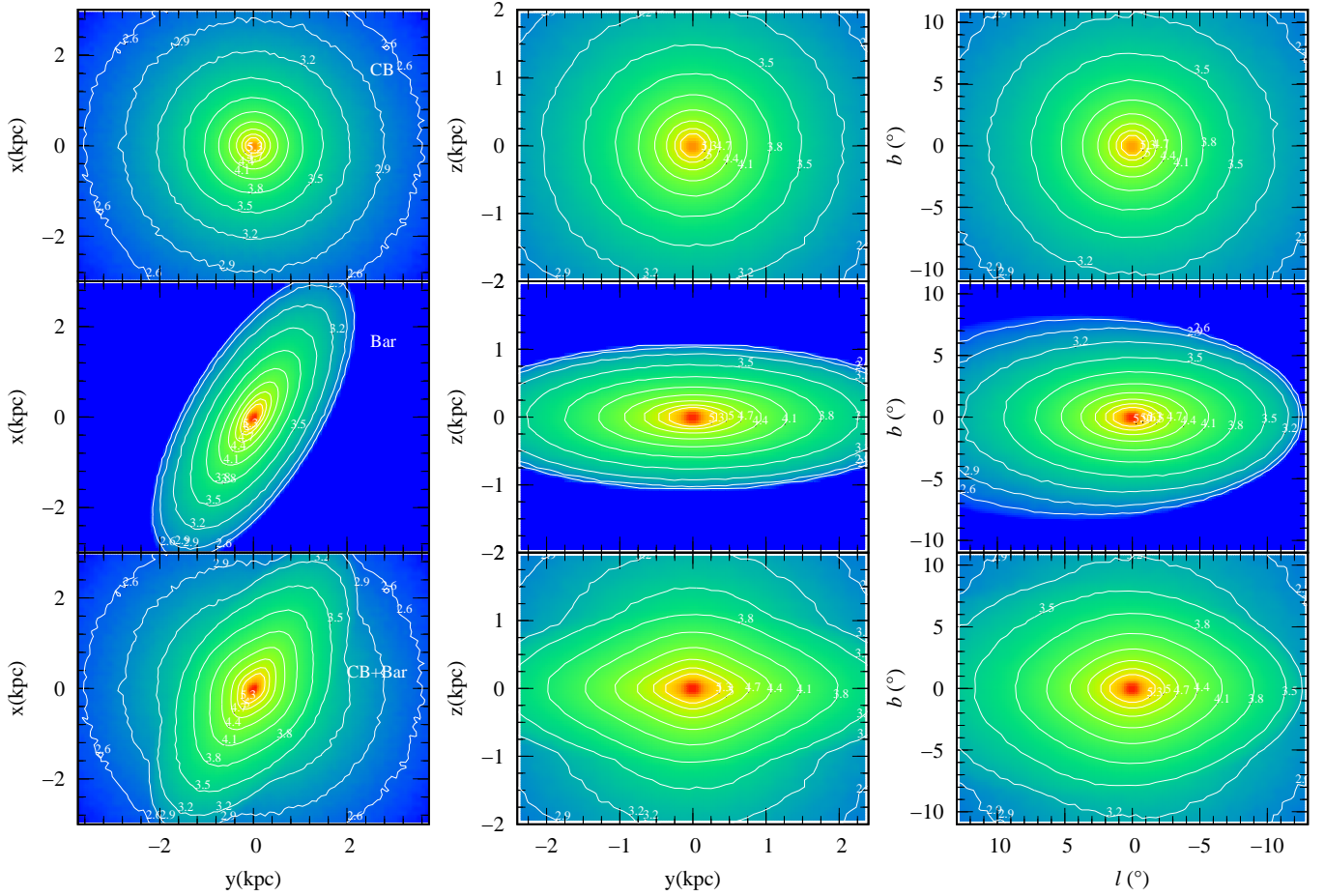}
\caption{
Density map of the two component bulge model.
Left panels are face-on views of the classical bulge, the bar, and the sum of the two, respectively, from top to bottom panels.
Middle panels are the same models but for the side-on views.
Right panels are perspective views from an angle of the Sun.
The numbers given along the iso-density contour lines are arbitrary star numbers within 0.01~kpc$^2$ (left and middle panels) and 0.25~deg$^2$ (right panels) in log scale.
They are only relevant for the relative density scales between the classical bulge and the bar components.
Total number of stars used in the simulatation is $5.5\times10^7$.
\label{fig10}}
\end{figure*}

\begin{figure*}
\includegraphics[angle=-90,scale=0.9]{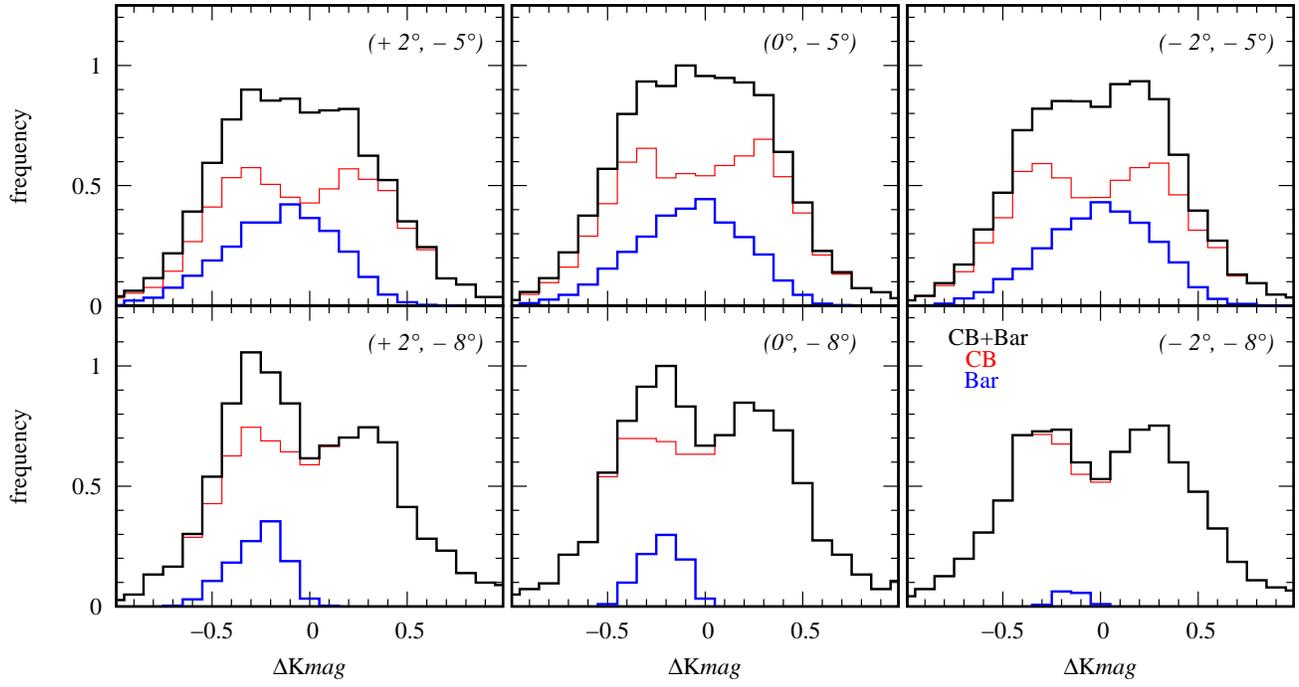}
\caption{
The longitude dependence of the RC luminosity function at $b=-5$ and $-8\degr$ fields.
The bimodal histograms in red are for the classical bulge component, while the unimodal histograms in blue are for the bar component.
The effects of the stellar density and the fractional change between the two components as predicted in Figure~\ref{fig10} are included.
Black histograms are the sum of the two components.
These models show good agreements with the observed histograms in Figure~2 of \citet{gon15}.
\label{fig11}}
\end{figure*}

\begin{figure}
\epsscale{1.0}
\plotone{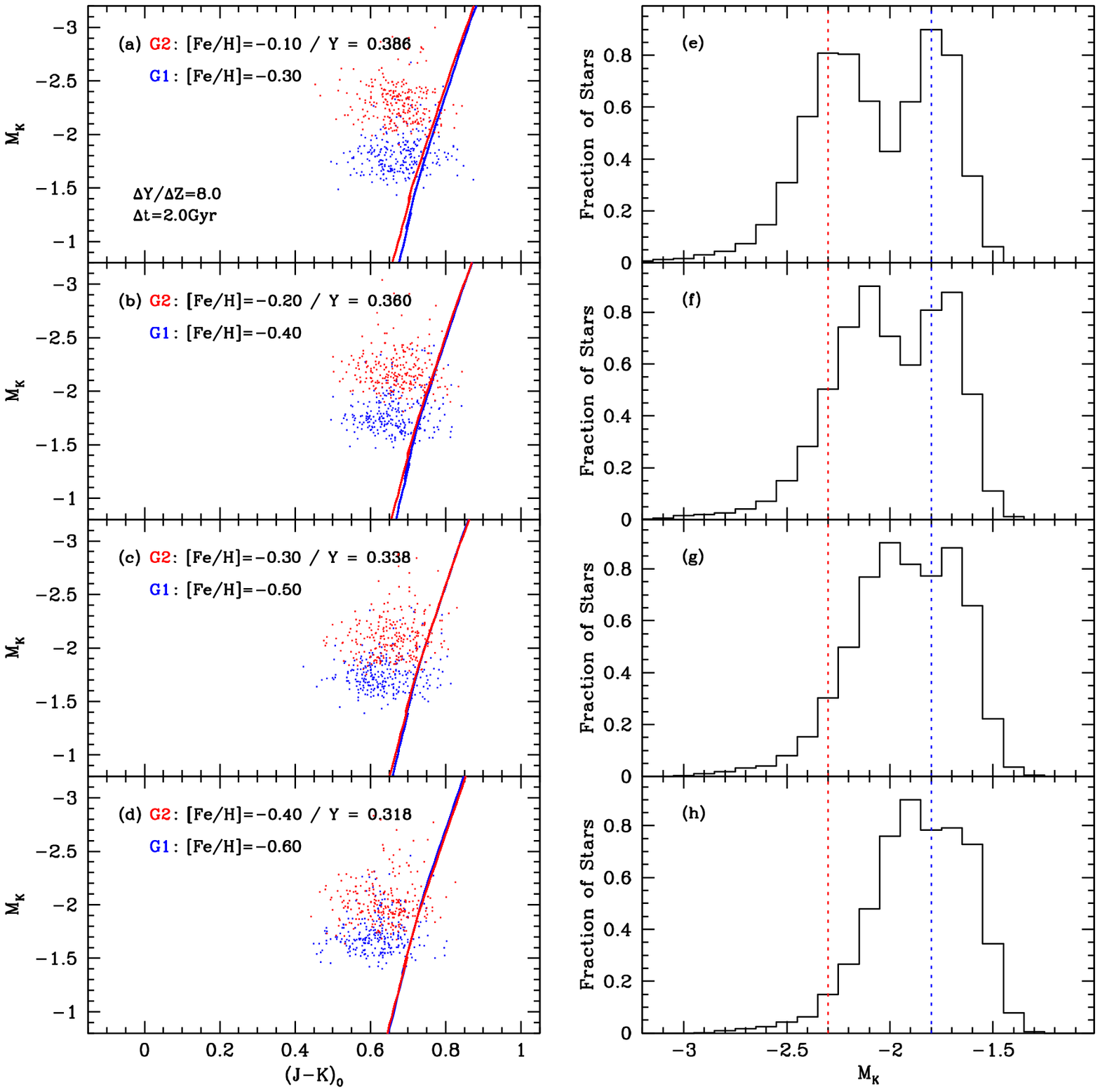}
\caption{
Effect of the metallicity variation on the RC models for the $(l,b)= (0\degr,-8\degr)$ to $(-4\degr,-8\degr)$ fields,
in $(J-K,K)$ CMDs and $K$-band luminosity functions. The metallicity scale is from the bulge metallicity map by \citet{gon13},
i.e., $<$[Fe/H]$> = -0.2$ to $-0.5$ from top to bottom panels. Here, we adopted $\Delta Y/\Delta Z = 8.0$ for G2, and assumed 11 and 9 Gyrs for
the ages of G1 and G2, respectively. The magnitude and color errors ($\sigma_{K} \approx 0.093$, $\sigma_{J-K} \approx 0.066$, and
$\sigma_I \approx 0.074$, $\sigma_{V-I} \approx 0.052$)
were included in our simulations to reflect basic photometric errors, [Fe/H] spreads, differential reddening,
and optical depth effect. Note that the double RC is merged into one by only a small change in metallicity $\Delta$[Fe/H] = 0.2 $-$ 0.3 dex.
The vertical dotted lines represent two peaks of the RC luminosity function for the models in panels~(a) and (e).
\label{fig12}}
\end{figure}

\begin{figure}
\epsscale{1.0}
\plotone{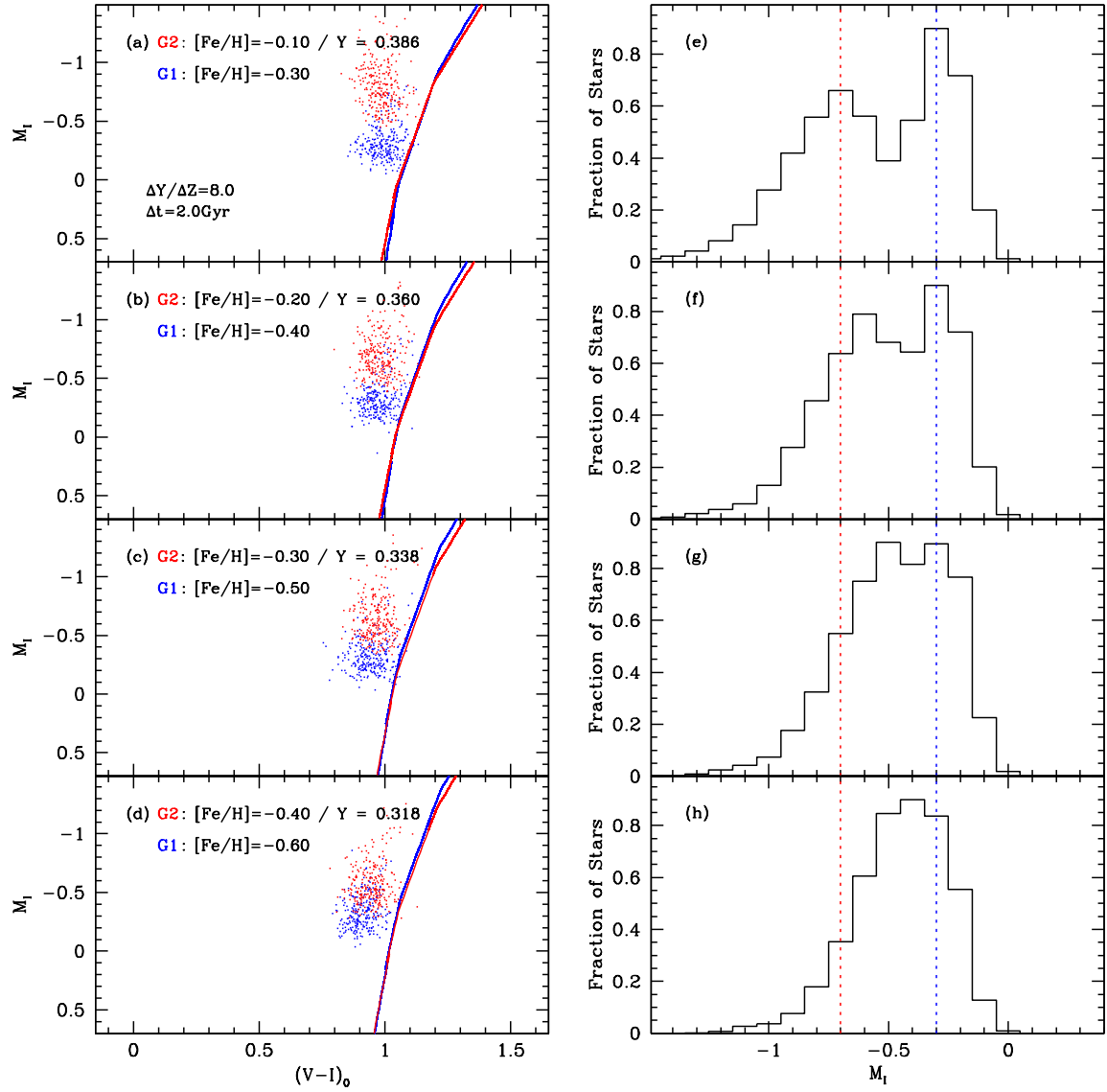}
\caption{
Same as Figure~\ref{fig12}, but for $(V-I,I)$ CMDs and $I$-band luminosity functions.
\label{fig13}}
\end{figure}

In the relatively metal-poor regime ([Fe/H] $\approx -1.1$) of the Milky Way bulge, \citet{pie15} discovered two sequences of RR Lyrae stars on the period-amplitude diagram.
Recently, \citet{lee16} suggested a common origin for this phenomenon and the double RC observed in the metal-rich regime of the bulge.
They have shown that the period-shift between these two populations of RR Lyrae stars is due to a small difference in helium abundance ($\Delta$Y $\approx 0.01$)
between the first- and second-generation stars. As discussed above, this is expected because $\Delta$Y between G2 and G1 would be
an order of magnitude smaller amount at this low metallicity regime compared to the metal-rich bulge where the double RC is observed.
Therefore, it appears quite possible that the two populations of RR Lyrae stars and the double RC observed in the Milky Way bulge might be different
manifestations of the same multiple population phenomenon in the metal-poor and metal-rich regimes respectively. It is important to note that the double RC
originated from multiple population phenomenon is also observed in a metal-rich bulge GC Terzan~5 \citep{fer09}. In Figure~\ref{fig8}, we estimate
a helium difference of $\Delta$Y $\approx 0.07$ between the two RCs in Terzan~5, which is roughly consistent with the value expected from
a $\Delta Y/\Delta Z = 6.0$ trend at the metallicity of this GC ([Fe/H] = $-$0.2; \citealt{har96}). At lower metallicity, $\Delta$Y between G2 and G1
would be smaller and consequently the magnitude separation between the two RCs would be also narrower, or the two RCs could be merged into one
with a vertically extended feature (see Figure~\ref{fig3}). Interestingly, the RC part of the HB in NGC~6441, another bulge GC with [Fe/H] = $-$0.5
\citep{pio02,har96}, shows just this feature, and \citet[][see their Fig. 6]{cal07} indeed estimate a smaller $\Delta$Y between G2 and G1
($\Delta$Y $\approx 0.05$) from the RC luminosity function.

As stressed in \citetalias{lee15}, Gaia trigonometric parallax distances can soon provide a crucial test as to the origin of the double RC observed in the
Milky Way bulge. This will in turn reveal whether a non-negligible fraction of the classical bulge population is embedded in a bar structure of the Milky Way or not.
Since the stellar populations in the classical bulge component of the Milky Way is considered as the nearest example of those in elliptical galaxies,
this will also severely impact our understanding of stellar populations and formation of early-type galaxies.

\acknowledgments
We thank the referee for a number of helpful suggestions. Y.-W.L. acknowledges support from the National Research Foundation of Korea to the Center for Galaxy Evolution Research.

\clearpage

\clearpage

\clearpage

\clearpage

\clearpage

\clearpage

\clearpage

\clearpage

\clearpage

\clearpage

\clearpage

\clearpage

\clearpage

\clearpage

\end{document}